\documentclass[%
 aip,
 amsmath,amssymb,
 reprint,%
]{revtex4-1}
\usepackage{orcidlink}
\usepackage{graphicx}% Include figure files
\usepackage{dcolumn}% Align table columns on decimal point
\usepackage{bm}% bold math
\usepackage[version=4]{mhchem}
\usepackage[utf8]{inputenc}
\usepackage[T1]{fontenc}
\usepackage{mathptmx}
\usepackage{etoolbox}

\makeatletter
\def\@email#1#2{%
 \endgroup
 \patchcmd{\titleblock@produce}
  {\frontmatter@RRAPformat}
  {\frontmatter@RRAPformat{\produce@RRAP{*#1\href{mailto:#2}{#2}}}\frontmatter@RRAPformat}
  {}{}
}%
\makeatother
\begin{document}

\preprint{AIP/123-QED}

\title{A versatile setup for symmetry-resolved  ultrafast dynamics of quantum materials}

\author{Khalid M. Siddiqui \orcidlink{0000-0001-6640-2406}
}
\email{khalid.siddiqui@phys.au.dk}
\affiliation{Department of Physics and Astronomy, Aarhus University, 8000 Aarhus, Denmark.}
 
\author{Hanna Strojecka \orcidlink{0009-0002-8992-3620}
}
\affiliation{Department of Physics and Astronomy, Aarhus University, 8000 Aarhus, Denmark.}
\affiliation{Univ Rennes, CNRS, Institut de Physique de Rennes – UMR 6251, 35000 Rennes, France.}

\author{Thomas H. Meyland \orcidlink{0009-0005-8344-7469}
}
\affiliation{Department of Physics and Astronomy, Aarhus University, 8000 Aarhus, Denmark.}

\author{Nitesh Khatiwada \orcidlink{0009-0004-2584-0054}
}
\affiliation{Department of Physics and Astronomy, Aarhus University, 8000 Aarhus, Denmark.}
\affiliation{Department of Physics, University of Erlangen-Nürnberg, Erlangen, Germany.}

\author{Nikolaj Klinkby \orcidlink{0009-0008-0317-116X}
}
\affiliation{Department of Physics and Astronomy, Aarhus University, 8000 Aarhus, Denmark.}

\author{Daniel Perez-Salinas \orcidlink{0000-0002-4641-8387}
}
\affiliation{Department of Physics and Astronomy, Aarhus University, 8000 Aarhus, Denmark.}
\affiliation{ALBA Synchrotron Light Source, Cerdanyola del Valles, 08290 Barcelona, Spain.}

\author{Simon E. Wall \orcidlink{0000-0002-6136-0224}}
\email{simon.wall@phys.au.dk}
\affiliation{Department of Physics and Astronomy, Aarhus University, 8000 Aarhus, Denmark.}

\date{\today}% It is always \today, today,

\begin{abstract}
Correlated phenomena occur in quantum materials because of the delicate interplay between internal degrees of freedom, leading to multiple symmetry-broken quantum phases. Resolving the structure of these phases is a key challenge, often requiring facilities equipped with x-ray free-electron lasers and electron sources that may not be readily accessible to the average user. Table-top sources that offer alternative means are therefore needed. In this work, we present an all-optical, table-top setup that enables symmetry-resolved ultrafast studies of quantum materials using linear and nonlinear spectroscopies. We demonstrate the versatility of the setup with chosen examples that underscore the importance of tracking symmetries and showcase the strengths of the setup, which offers a large tunable parameter space. 

\end{abstract}

\maketitle

\section{Introduction}
\label{section:intro}
Ultrafast control of materials leading to novel states with exotic properties has become a central goal in condensed matter physics, offering potentially promising technological applications\cite{Basov2017}. Quantum materials are especially amenable to such protocols due to their intrinsic susceptibility to external stimuli (e.g., doping, pressure, temperature, etc.), which lead to new and complex phase diagrams \cite{Tokura2017,delaTorre2021,Ravnik2021}. Indeed, using ultrashort laser pulses across a broad spectrum\textemdash from near-IR to THz\textemdash novel phases have been reported, including light-induced superconductivity \cite{Fausti2011}, hidden metastable phases \cite{Stojchevska2014} and induced chirality\cite{Zeng2025}, among others. These findings have earned quantum materials the reputation as a fertile source for emerging non-equilibrium phases \cite{Zhang2014}.  

The ability to control emergent phases is inherently linked to their robust detection and characterization. A hallmark of structural phase transitions in quantum materials is symmetry-breaking; thus, directly tracking symmetry-related quantities connected to the order parameter can yield crucial insights into the underlying properties and mechanism\cite{Tzschaschel2019,Wandel2022}. Optical pump-probe techniques\textemdash in which a laser pulse creates a transient state, and another time-delayed optical probe is used to interrogate its properties\textemdash have long served as a powerful method to monitor ultrafast symmetry changes, e.g. in lattice using coherent phonon signatures \cite{Wall2012,Stojchevska2014}. However, data interpretation is not always straightforward and can be impeded by the presence of a complex electronic response. On the other hand, ultrafast diffraction techniques based on x-ray free-electron lasers (XFEL)\cite{Mitrano2024,Wall2016} and ultrabright electron sources\cite{Filippetto2022,Siddiqui2023} offer unique perspectives as direct probes of the lattice and its interplay with other subsystems (electronic, spin, orbital). However, these methods are costly, technically demanding, and require specialized expertise, making them inaccessible to many users.

Recognizing this, several groups have developed new table-top techniques to elucidate symmetry-breaking processes in materials, statically and on ultrafast timescales. Polarization-sensitive optical spectroscopies exhibit high sensitivity to the point group of materials \cite{ Denev2011} and have proven especially powerful for investigating electronic, magnetic, rotational, and nematic order dynamics \cite{Patz2014,Toda2014,PerezSalinas2022}, including nonlinear probes capable of resolving more subtle symmetry-breaking phenomena (e.g. involving inversion and mirror symmetries) as well as hidden quantum orders \cite{Padmanabhan2018,Fichera2020}. The general aim is to measure the polarization dependence of the linear or nonlinear optical properties of the material. Two schemes have been applied: the first entails rotating the sample around a selected crystal axis and recording the reflected or transmitted light of fixed polarization \cite{Tom1983,Glinka2015}; the second involves sampling the input and output light polarizations using an input-polarizer output-polarization-analyzer combination while the sample is kept stationary \cite{torchinsky2014low,Ali2021,Morey2025,Zhang2024}. From the resulting rotational anisotropy (RA) patterns insight into the intrinsic point group symmetries can be obtained \cite{Harter2015,torchinsky2016rotational,Sala2016}. Thus, RA spectroscopy is an effective complementary probe to conventional structural methods, such as x-ray diffraction (XRD).

A key challenge in both linear and nonlinear RA spectroscopies is avoiding drifts in the laser intensity during measurements, which can obscure subtle polarization-dependent changes. This issue is especially critical in ultrafast measurements, which add an additional dimension to the data collection and can drastically increase the measurement time. Therefore, it is crucial to complete a full anisotropy scan before any substantial laser drift occurs. As it is not possible to rapidly spin the sample in cryogenic conditions, the second scheme has been utilized which involves rotating the polarization of the incoming light. Hsieh and co-workers developed such a second harmonic generation rotational anisotropy (SHG-RA) apparatus for symmetry-resolved optical studies, incorporating a custom-built phase mask (PM) co-spinning  with the input polarizer, thereby sweeping the scattering plane about a fixed point on the sample. This mapped the polarization state onto a specific reflection angle, collected on a position-sensitive charge-coupled device (CCD) \cite{Harter2015}. This approach works well for static characterizations, but the slow readout of the CCD makes it less ideal and time-consuming for pump-probe measurements where signals must additionally be sorted into pumped and unpumped bins \cite{Shan2021}. The use of custom diffractive optics (e.g., PM), moreover, requires tailoring them to a specific wavelength, making spectroscopy experiments extremely demanding.

Lu \textit{et al}. used a modified approach in which they employed fast rotating, reflective optics permitting wavelength-dependent experiments \cite{Lu2019}. Additionally, a lens was used to focus the varying output light to the same point, thereby allowing a more standardized detection using a photomultiplier tube (PMT) resulting in rapid digitization of the signal. These setups have been successfully employed to probe hidden orders in spin-orbit-coupled iridates \cite{LZhao2015}, detect mirror symmetry-breaking across the pseudogap phase of cuprates\cite{Zhao2016} and investigate structural changes under extreme pressure \cite{Li2022}.

Most standard RA setups, such as those mentioned above, employ oblique incidence geometry, which is generally preferred as it enables probing all crystallographic axes via light polarization \cite{Denev2011}. However, this geometry can potentially complicate analysis, since uniquely determining all tensor elements often requires measurements at multiple incident angles. In systems where the angle is fixed by a focusing objective, as in the setups above, adjusting it is no longer straightforward. This is especially true in cryogenic environments, where tilting is restricted by physical constraints imposed by the cryostat, and the need to maintain near-normal incidence to cryostat windows to avoid polarization effects. Moreover, changing the angle can alter both sample reflectivity and the pump beam footprint with the former being particularly difficult to model if the pump modifies reflectivity in a non-trivial way. As a result, streamlined strategies are needed to overcome these challenges and enable efficient pump-probe operation without introducing further technical complexity.

In this article, we present a new versatile instrument for symmetry-resolved ultrafast dynamics of quantum materials via rotational anisotropy spectroscopy that enables a diverse set of tunable parameters in a simple and cost-effective implementation. The basis of our design is a near-normal incident light on the sample double-passing through a superachromatic half-wave plate with ultrabroadband coverage (visible to the entire near-infrared), which is the sole rotating element in our setup. This distinguishing feature reduces alignment issues associated with multiple rotating elements while minimizing the overall setup footprint and enabling experiments involving fundamental and frequency-doubled light. Unlike the setups above, the normal-incidence approach is a more restricted, in the sense that it is ideal for studying in-plane symmetry-breaking, and minimizes out-of-plane axis or surface effects. Probing other crystallographic axes still remain accessible, however, through appropriately polished single crystalline bulk samples of quantum materials. As will be shown, this approach achieves high-fidelity results that enables resolving symmetries in quantum materials with high signal-to-noise (S/N) ratios using both linear and nonlinear spectroscopies.

The paper is divided into four sections. Section II begins by providing a brief overview of the theoretical formalism behind rotational anisotropy experiments, followed by a detailed description of the RA setup. It also includes a discussion on setup characterization and provides some key performance metrics. Section III presents selected scientific examples of symmetry changes in quantum materials studied through linear and nonlinear RA spectroscopies. This section concludes with a straightforward approach proposed for detecting extremely weak signals encountered in SHG-RA experiments that can be easily combined with the linear RA setup. Finally, in Section IV, the paper concludes with a summary and a future outlook.

\section{Rotation Anisotropy Spectroscopy}
\label{section:RA setup}
\subsection{Theoretical Methodology}
For many quantum materials, the order parameter can be associated with the loss of rotational symmetry (e.g., \ce{C4}), leading to the emergence of optical birefringence\cite{PerezSalinas2022,Toda2014,Lubashevsky2014} as the system transitions from a high to low symmetry state. Therefore, the degree of symmetry-breaking can be inferred from the material's optical anisotropic response and in this section, we describe the theoretical framework that underpins rotational anisotropy experiments designed to probe birefringent behavior. 

For a given material, we can define a matrix, $r$, given as,

\begin{align}
r =  \begin{pmatrix}
r_1e^{i\phi_1} & 0  \\
 0& r_2e^{i\phi_2} \\
\end{pmatrix},
\end{align}

where $r_{1,2}$ denote the in-plane reflection coefficients, and $\phi_{1,2}$ are the associated phase shifts. We introduce $\rho = \frac{r_2}{r_1}$  and $\chi = \phi_2 - \phi_1$, which quantify the birefringence and ellipticity (imparted by the sample), respectively. Based on these definitions, an expression for an effective matrix, $r_{eff}$ can be derived using Jones algebra\cite{Pistoni1995} for the case where the light goes through a polarizer and then double-passes through a rotating half-wave plate (HWP) before and after reflection from the sample. This gives,
 
\begin{align}
r_{eff} =  \alpha\begin{pmatrix}
\cos^2\theta + \rho e^{i\chi}\sin^2\theta & \frac{1}{2}\sin2\theta(1-\rho^{i\chi})  \\
 \frac{1}{2}\sin2\theta(1-\rho^{i\chi})& \cos^2\theta + \rho e^{i\chi}\sin^2\theta \\
\end{pmatrix},
\end{align}

where $\alpha = r_1e^{i\phi_1}$ and $\theta$ is the angle made by the input electric field relative to the crystal axis. For a horizontally polarized beam, i.e. $
\vec{E} = \begin{bmatrix}
           1 \\
           0 \\
           \end{bmatrix} $
incident on the sample, which then passes through the polarizer a second time, one obtains the expression for intensity, $I_{sig}$, which varies with $\theta$ as,

\begin{equation}   
\begin{aligned}
I_{sig}(\theta) = & \frac{r_{1}^{2}}{8} \left( 4(1-\rho^2)\cos2\theta + (1 + \rho^2 - 2\rho\cos\chi)\cos4\theta \right. \\
   & \left. + 6 + 2\rho\cos\chi \right).
\end{aligned}
\end{equation}

Note that the use of same polarization optics ensures that the detected beam is perfectly parallel polarized. In the examples presented in section IIIA of this paper, we fit the data as a function of the half-wave plate angle, $\Theta$, (which controls the polarization as $\theta=2\Theta$) using the expression, 

\begin{equation}
    I_{sig}(\Theta) = I_{DC} +  I_4\cos(4\Theta + \Psi_4) +I_8\cos(8\Theta + \Psi_8),  
    \label{Eq:1}
\end{equation}
\begin{figure*}[t!]
    \centering
    \includegraphics[width=0.9\textwidth]{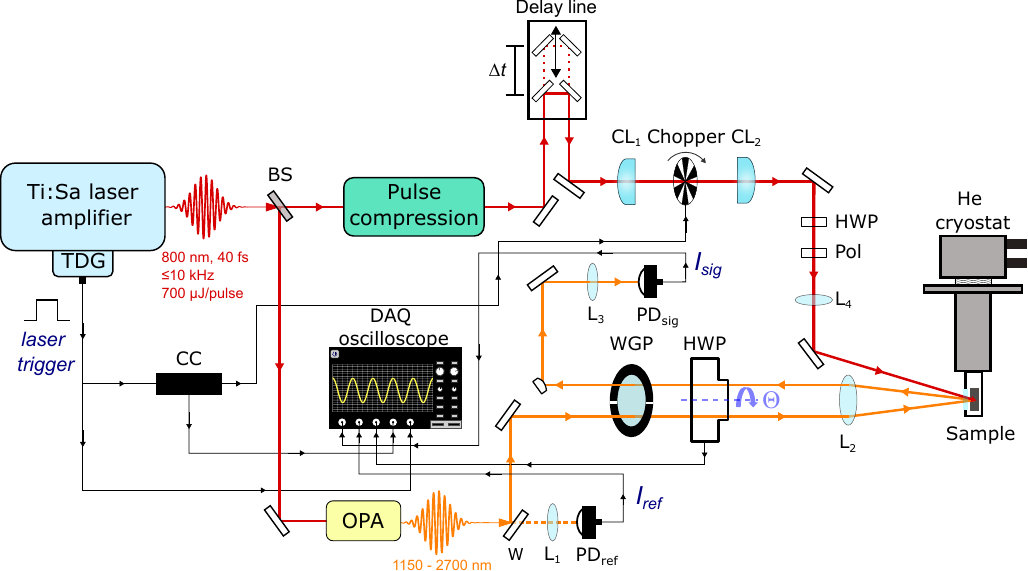}
    \caption{Schematic of linear and non-linear rotational anisotropy setup. BS: beam-splitter; TDG: timing and delay generator; CC: chopper-controller; OPA: optical parametric amplifier;  $\mathrm{PD_{sig/ref}}$: signal/reference photodiode; L: lens; W: wedge prism;  CL: cylindrical lens; WGP: broadband wire-grid polariser; HWP: broadband half-wave plate. Pol: polarizer; DAQ: data acquisition. }
    \label{Fig1:RASetup}
\end{figure*}

where $I_{DC}$ relates to the isotropic contribution, $I_4 (I_8)$ are four-fold (eight-fold) anisotropic terms, and $\Psi_{4/8}$ denote their respective phase angles. $I_8$ can originate from orthogonal channels, e.g. $I^{PS}_{sig}, I^{SP}_{sig}$, where supercripts \textit{P} and \textit{S} denote parallel and perpendicular polarization states, respectively, but its contribution to $I^{PP}/I^{SS}$ signal is expected to be very small. Additional terms can be included in Eq.~(\ref{Eq:1}) to account for experimental artifacts, such as light backscattering from the rotating half-wave plate reaching the detector and misalignment of the incident polarization relative to the crystal axis. Note that the effects of mechanical strain are not considered in this analysis. 

Using the information above, we can derive expressions for the birefringence ratio, $\rho$, the order parameter, $\eta^2$ (defined as proportional to $r_1-r_2$), and reflectivity, $r_1$ as follows (see Ref.~\!\citenum{DaniThesis} for details): 

\begin{align}
    &r^2_{1} = \frac{4I_8+4I_{DC}+I_4}{4},\\
   & \rho^2 = \frac{4(I_8 + I_{DC}) -7I_4}{I_4+4(I_8+I_{DC})},\\
  &  \eta^2 = \frac{1-\rho}{1+\rho},
\end{align}

where the order parameter, $\eta^2$ has been normalized to lie in $\pm$1. Note that in our analysis when $\rho =0$, it denotes the high symmetry state. Therefore, we can obtain direct insights into symmetry changes and sample electronic response from rotational anisotropy experiments, which motivates the setup described in the next section.

\subsection{The Setup}
A schematic of the rotational anisotropy setup is provided in Fig.~\ref{Fig1:RASetup}. The laser system comprising a commercial titanium-doped sapphire (Ti:Sa) oscillator and regenerative amplifier produces a train of near-infrared pulses at 800 nm with a maximum pulse energy of 700 \textmu J and a tunable repetition rate of up to 10 kHz. The fundamental output of the amplifier is divided into two arms using a 90:10 ratio beam splitter (BS). The reflected portion carrying 90\% of the total energy pumps a commercial collinear optical parametric amplifier (OPA), in which a white-light-supercontinuum-seeded Beta-Barium Borate (BBO) crystal is pumped by 800 nm pulses and amplifies a broad range of wavelengths in the near-infrared, producing signal (1.15--1.6~\textmu m) and idler (1.6--2.7~\textmu m) photons. The output of the OPA mainly serves as a probe in our setup but can be used as a pump on demand. At the OPA exit, dichroic filters are placed to separate the signal and idler beams and their corresponding paths to the sample have been matched to permit convenient interchanging. 

Following the OPA, the probe beam is split by a wedge (W), and the reflected portion is directed by a pick-off mirror towards the sample. On the way, it first encounters a wire-grid linear polarizer (WGP), which fixes the initial polarization state. The beam then goes through an ultrabroadband ($\lambda = 0.6 -2.7$~\textmu m, phase retardance error $<1\%$ in this range), superachromatic half-wave plate, i.e. HWP, that is mounted onto a freely-spinning rotation stage (rate tunable up to 10 Hz), creating multiple polarization states. As stated previously, we use the ultrabroadband coverage provided by the HWP for both linear and nonlinear optics experiments. After passing through the HWP, the probe beam is imaged onto the sample at near-normal incidence by a 25 cm focal length lens (L$_2$) to a full-width-at-half-maximum (FWHM) spot size of $\approx$ 60 \textmu m in diameter. We estimate the angle of incidence to be approximately $2.3^\circ$, corresponding to a suppression of the out-of-plane signal by about a factor of 25.

Bulk crystals or thin film samples are mounted inside a closed-cycle helium (He) cryostat fitted with vibration isolation bellows. We use a 2 mm thick non-birefringent IR-silica as the cryostat entrance window material to reduce the contribution of birefringence from sources other than the sample. The sample temperature is controlled via a proportional-integral-differential (PID) loop between 7 K and 420 K, allowing measurements of a wide range of materials from conventional and high-temperature superconductors\cite{Xu2019,Tian2016}, charge density waves (CDW) \cite{Sayers2022}, Mott insulators \cite{Johnson2022}, magnetic materials \cite{Ergeen2023} and more. 

In our chosen geometry, the probe is retroreflected by the sample, being horizontally displaced with respect to the incoming beam. It passes through the same optics as on the incident path, i.e. Sample $\rightarrow$Lens$\rightarrow$ HWP$\rightarrow$WGP. As mentioned earlier, the combination of near-normal incidence and double-pass through the same optics simplifies alignment, removes errors due to intrinsic birefringence (see later) and, reduces setup footprint. We find that small deviations in sample flatness do not affect the alignment, and minor angular misalignments can be manually corrected using routing mirrors or slight tilt adjustments of the sample holder. The light transmitted through the WGP on the return path is picked up by a D-shaped mirror and focused by $L_3$ onto a germanium (Ge) or indium gallium arsenide (InGaAs) photodiode ($\mathrm{PD_{sig}}$), depending on the probe wavelength. The use of $L_3$ ensures efficient signal collection on the detector in nonlinear experiments, similar to  many RA setups, even though wavelength changes result in a beam that is not perfectly collimated. The $\mathrm{PD_{sig}}$ measures the intensity of the signal reflected from the sample ($I_{sig}$), while a small fraction of the light ($I_{ref}$), transmitted through the wedge in the probe arm, is simultaneously focused by L$_1$ onto a reference photodiode ($\mathrm{PD_{ref}}$). Both signals, $I_{sig}$ and $I_{ref}$, are sent to the data acquisition (DAQ) system for recording, with the latter used for post-normalization of the signal to minimize the effects of probe fluctuations and increase the signal-to-noise ratio.

The remaining 10\% of the light at 800 nm transmitted through the BS is reserved for photoexcitation. The pump beam is first compressed by a set of chirped mirrors\textemdash which impose a negative group delay dispersion\textemdash before being sent into a computer-controlled optical delay line providing up to 800 ps in maximum pump-probe delay. The beam subsequently travels: (\textbf{1}) through a mechanical chopper enclosed within a telescope constructed using cylindrical lenses (CL$_{1,2}$). The choice of cylindrical focusing enables chopping at 5 kHz using standard optical chopper blades, while the resulting line focus ensures that the peak intensity remains well below the threshold for air filamentation; (\textbf{2}) a 800 nm HWP and polarizer (Pol) pair for precise power control and finally, (\textbf{3}) focused onto the sample by a 50 cm lens, (L$_4$) to a spot size of $\approx$~170 \textmu m FWHM at an incident angle of about 4 degrees with respect to the probe beam.   

A flip mirror is placed in front of the sample, before the cryostat, to direct the pump and probe beams to an 8-bit CMOS camera\textemdash positioned at the focal point of the probe\textemdash with 1.6 \textmu m pixel size for beam size characterization. The spatial overlap between the pump and probe beams is first established here before being optimized at the sample plane. It is also here that the time-resolution characterizations are carried out (see below). The repetition rate, polarizations, pump fluence, pump/probe colors, and sample temperature can all be changed and controlled, which allows tailoring the setup to study a range of materials under different settings.

\subsubsection*{Data collection}
We now describe the data collection procedure for static and time-resolved RA experiments. For each measurement type, the following quantities are recorded: intensities of the signal and reference diodes ($I_{sig}$ and $I_{ref}$), the rotation stage zero-crossing, the chopper state, delay position and fluence values. The latter three are typically not required for static experiments, except when performing pump-probe measurements at a fixed time delay. 
\begin{figure}[t!]
    \centering
    \includegraphics[width=8.65cm]{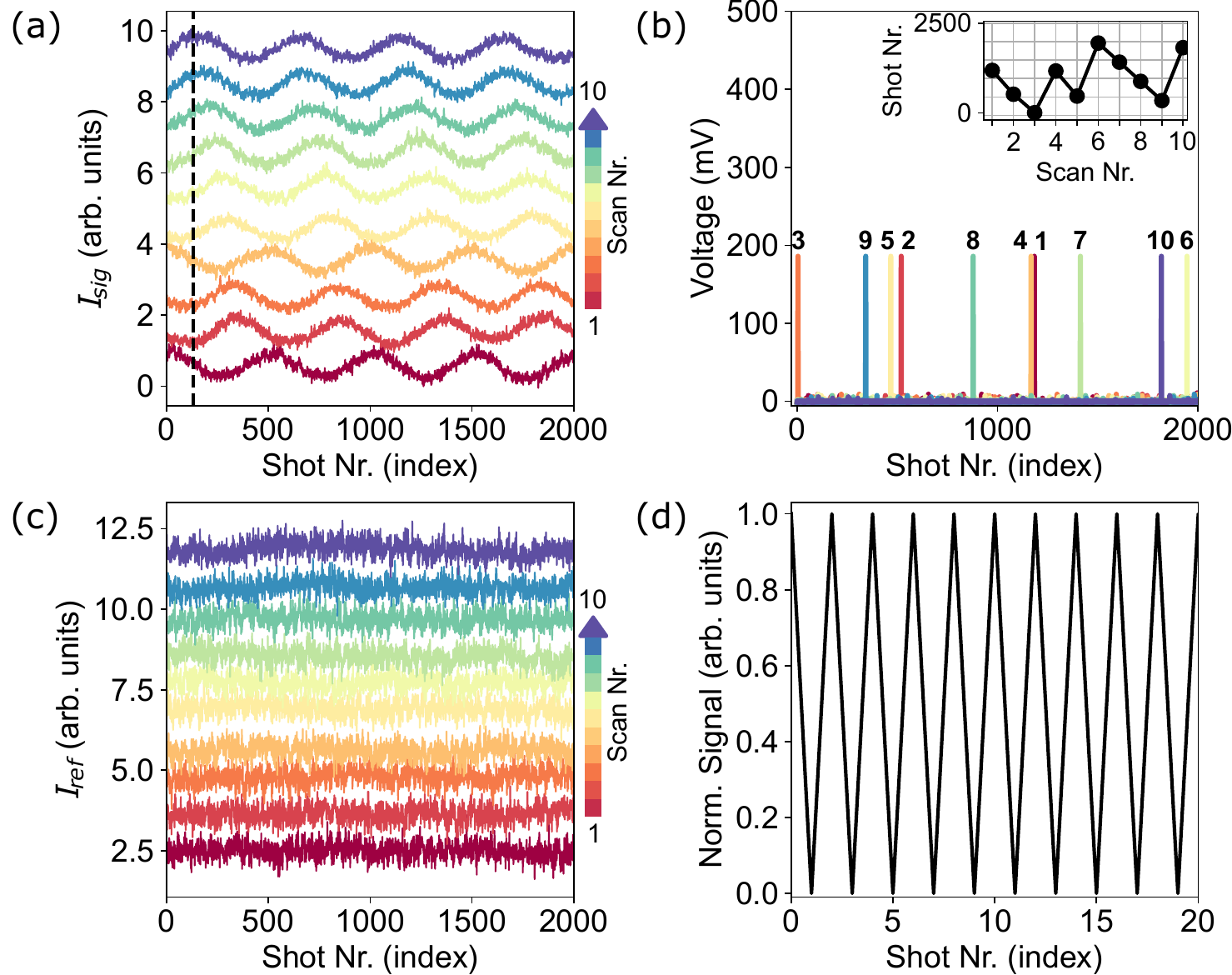}
    \caption{(a). Signal ($I_{sig}$) from a birefringent sample for ten consecutive scans as recorded by the DAQ system. Traces have random initial phases as denoted by the dashed line and need to be sorted before they can be averaged. The signals are vertically offset for clarity (b). The zero-crossing signal for each scan is shown as a vertical line. The label denotes the scan number out of ten measurements. Inset: locations of the zero-crossing which are used to shift and align the patterns before averaging (c). Signals from the reference diode, ($I_{ref}$) which are used for intensity normalization (d). Simulated normalized pump signal through the chopper demonstrating the alternating OFF and ON nature used for sorting the time-resolved data into respective Pump-OFF and Pump-ON bins. }
    \label{Fig2:DAQSignals}
\end{figure}

Figure~\ref{Fig2:DAQSignals} illustrates the signals captured by the four-channel, 1 GHz bandwidth DAQ oscilloscope. The scope is externally triggered using a transistor-transistor-logic (TTL) pulse derived from the laser's multi-channel timing delay generator (TDG), which initiates data acquisition. The DAQ system is equipped with a built-in analog-to-digital converter featuring high bandwidth and fast sampling rate, enabling the recording and processing of up to 5000 samples with a maximum rate of 20 GSa/s. It also includes processing modules, such as area trending, which we exploit to process the data on the fly. The signals from the diodes occupy two slots on the DAQ scope, whereas the other channels are reserved for signals from the chopper controller (CC) and the rotating HWP, respectively. 

Figure~\ref{Fig2:DAQSignals}a shows the acquired signals from ten successive scans of a birefringent sample, recorded with the laser repetition rate set to 10 kHz and rotation stage speed fixed at 5 Hz, i.e. 2000 laser shots made up a full rotation cycle. Each trace in Fig.~\ref{Fig2:DAQSignals}a represents a single-shot acquisition captured in just 50 milliseconds. As the rotation stage is freely spinning, the scan traces exhibit random initial phases as indicated by the dashed line. So, the first step is to sort the data according to the completion of a full rotation (i.e. zero-crossing). The stage controller was programmed to output an electronic pulse each time the stage underwent a full 360$^\circ$ rotation to locate the zero-crossing. Figure~\ref{Fig2:DAQSignals}b plots the moment (in terms of the shot number) from when the stage crossed the zero angle, shown as vertical lines, for each acquisition (scan numbers are labeled with an index). Once the zero angle has been determined, data sorting can be performed by shifting the traces to the respective zero-crossing, and waveplate angles can be assigned accordingly. The corresponding reference signals, measured by $\mathrm{PD_{ref}}$ and shown in Fig.~\ref{Fig2:DAQSignals}b, are used for normalization, as mentioned earlier. Normalized sorted data are plotted in real time during data acquisition and saved along with the raw patterns for later analysis.  

For pump-probe experiments, a mechanical chopper is spun at half the laser repetition rate, blocking every second pump pulse to create alternating Pump-ON and Pump-OFF sequences. The chopper controller (CC) outputs a train of pulses at the set chopping rate, as represented in Fig.~\ref{Fig2:DAQSignals}d. This signal is also sent to the DAQ scope and used to filter the data based on whether the probe pulse arrived in the time window when the chopper state was high (Pump-ON) or low (Pump-OFF). Moreover, a pump fluence calibration table is created pre-experiment and imported to define a single value or a list of fluences for scanning to control the excitation conditions. Finally, the delay information is dynamically recorded during scans by interfacing with the built-in encoder embedded in the optical delay line, which retrieves the absolute position of the delay stage at every step.

\subsubsection*{Setup characterization}
\begin{figure}[t!]
    \centering
    \includegraphics[width=8.5cm]{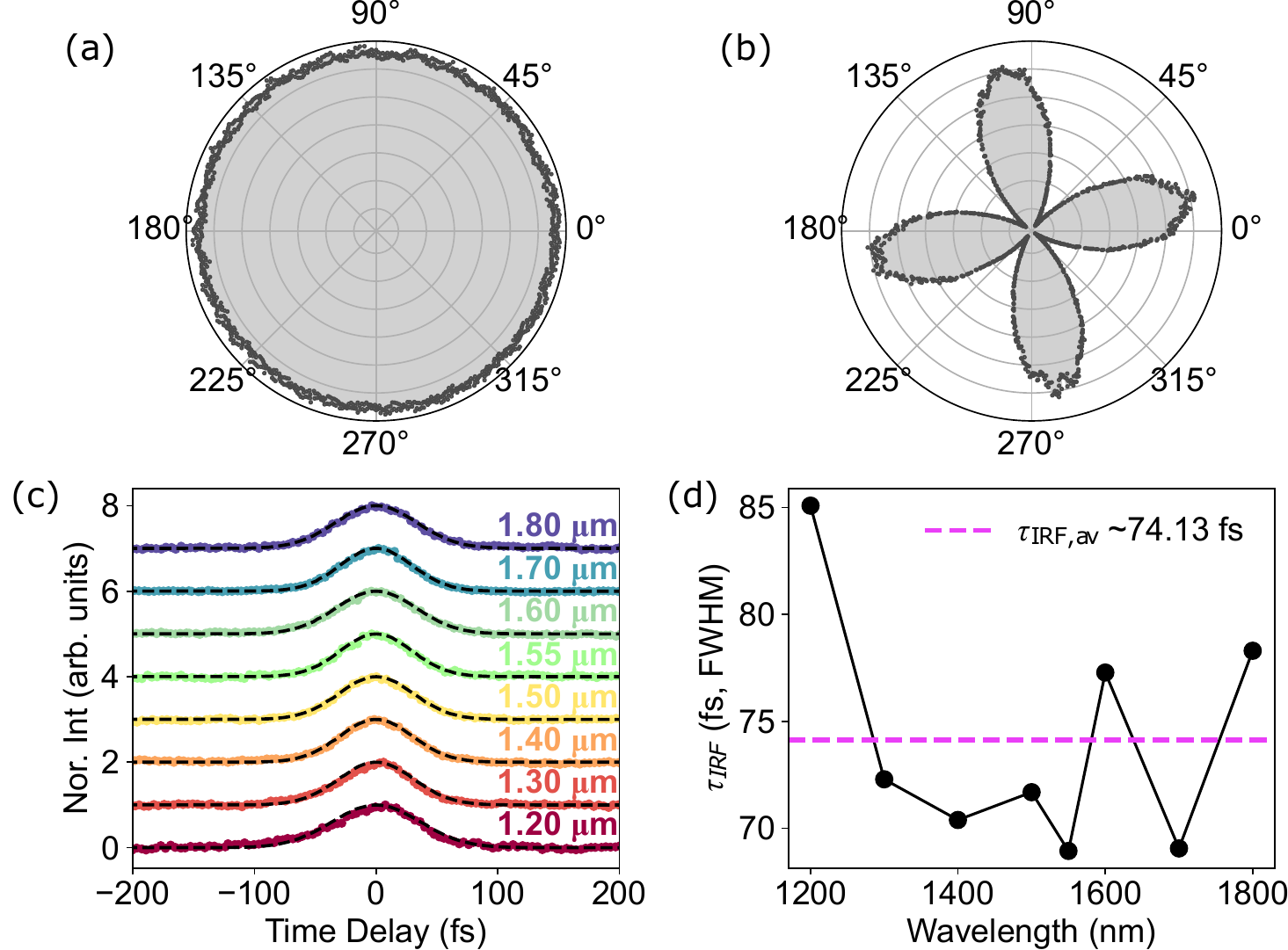}
    \caption{(a). Measured signal to characterize the degree of possible anisotropy solely due to optical elements carried out by placing a silver mirror in the sample plane and plotted in polar form. The circular pattern that emerges demonstrates very low anisotropic contribution and high isotropy. See text for more details (b). Measurement, repeated by placing a polarizer in the setup. (c). Measured traces resulting from cross-correlation between 800 nm pump and different probe wavelengths. The fits are given by the black dashed lines (d). Extracted FWHM width of the traces shown in panel (c) denoting the estimated time resolution as a function of probe wavelength. The average estimated value for instrument resolution of $\approx75$~fs is given as the dashed horizontal line.}
    \label{Fig3:SetupBenchmark}
\end{figure}

In this section, we discuss the characterization of the RA setup. We begin by examining the measured signals obtained within our setup to confirm that they are not affected by any extrinsic source of birefringence unrelated to the sample under investigation. These could originate from, e.g. mechanical strain in optical elements such as the focusing lens and the cryostat window, leading to stress-induced birefringence. To investigate this, we introduced a silver mirror in the sample plane with a high degree of isotropy and performed RA measurements. As the back-reflected beam traverses the same optical elements on its way to the detector as the incoming beam, the expectation is that this completely reverses any polarization induced from these optical elements save for the sample's intrinsic birefringence. Indeed, this was found to be the case, as the resulting signal (presented in Fig~\ref{Fig3:SetupBenchmark}a) was found to be insensitive to the polarization angle, producing a circular pattern when depicted as a polar plot. We note that similar results were obtained when the beam passed through a different lateral position on the cryostat window. Conversely, placing a polarizer in front of the mirror to mimic a birefringent sample produced the pattern in Fig.~\ref{Fig3:SetupBenchmark}b. These observations validate our design choice of a double-pass through the HWP. 

From these measurements, we can also estimate the sensitivity of our instrument to detect extremely low birefringence signals. To do this, we fit the above data with our birefringence model and determine the ratio between the anisotropic and isotropic terms, i.e. $I_4/I_{DC}$. We found the ratio to be $\approx1/400$, corresponding to about 0.25\% detectable polarization rotation, which can be considered the lower limit of anisotropy detection sensitivity in our setup. 

Next, we proceed to determine the temporal resolution achievable in our setup. Theoretically, the instrument response function can be given by $\tau_{IRF} = \sqrt{\tau_{\text{pr}}^2 + \tau_{\text{pu}}^2 + \tau_{\text{GVD}}^2 + \tau_{\text{jitter}}^2}$. Ignoring negligible group velocity dispersion (GVD) and jitter (i.e. $\tau_{\text{GVD}}$ \& $\tau_{\text{jitter}}$ = 0), $\tau_{IRF}$ can be seen as a convolution of the pump ($\tau_{\text{pu}}$) and probe ($\tau_{\text{pr}}$) pulse widths. We carried out cross-correlation measurements to determine our instrument's temporal response using 800 nm pump beam mixing in with the probe of different wavelengths inside of a 50~\textmu m thick BBO crystal. The signal generated in this experiment was filtered from the 800 nm beam using a combination of irises and spectral filters. The intensity of the signal was recorded using a Si photodiode. 
\begin{figure*}[t!]
    \centering
    \includegraphics[width=17.8cm]{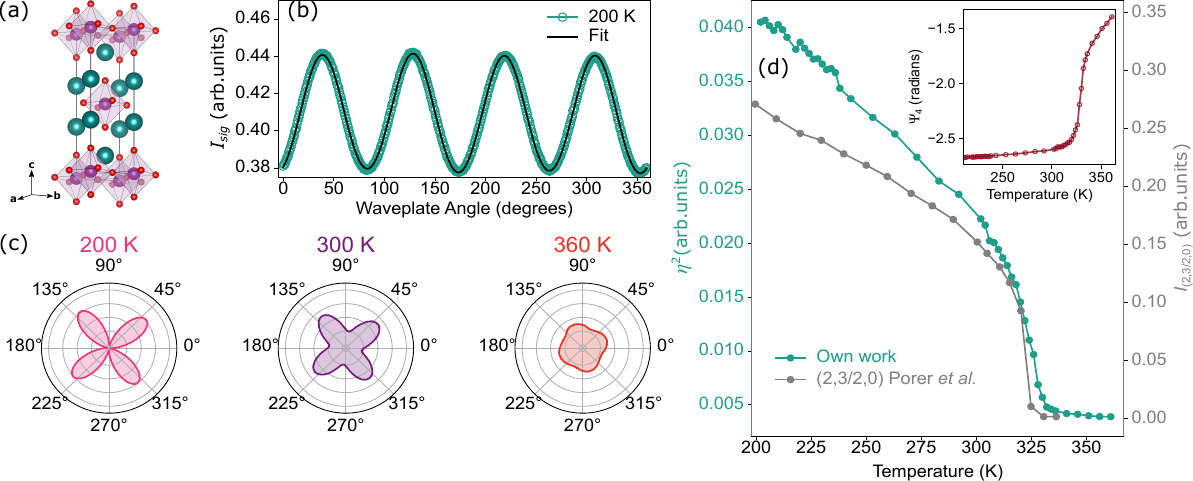}
    \caption{(a). Crystal structure of PCMO; Green denotes Ca/Pr atoms; purple: Mn, and red: oxygen atoms (b). Representative anisotropy and corresponding fit of measured birefringence at 200 K. (c). The change in anisotropy at selected temperatures represented in the form of polar plots. The amplitude reduces and a rotation of the pattern emerges. (d). Comparison between the measured temperature dependence of the order parameter from this work and that reported in the literature showing a good agreement between the two datasets. Data in gray is taken from Fig.~2 of Ref.~\!\citenum{Porer2020} with permission. Inset: Dependence of the $\Psi_4$ parameter which denotes the phase shift of the order parameter against temperature. }
    \label{Fig4:PCMO_TDep}
\end{figure*}
The results of these measurements are plotted in Fig.~\ref{Fig3:SetupBenchmark}c alongside fits assuming Gaussian profiles. Full-width-at-half-maximum widths extracted from the fits are plotted in Fig.~\ref{Fig3:SetupBenchmark}d. An average value of $\tau_{IRF} \approx 75$ fs (FWHM) is obtained for the time resolution of setup which is sufficient to track a variety of phenomena relevant in quantum materials, including electron-phonon coupling\cite{Pashkin2010}, CDW dynamics\cite{Sayers2022} and symmetry-breaking due to coherent motions (phonons\cite{Amuah2021}, magnons\cite{Rongione2023}) of up to 7 THz. 

\section{Results}
\subsection{Ultrafast symmetry-breaking in layered manganites} 

As a first demonstration of the capabilities of the setup, we investigated the symmetry-breaking phase transition in manganese oxide quantum materials. As a prototypical example of this class, we chose single crystals of \ce{Pr_{1.5}Ca_{0.5}MnO4} (henceforth denoted PCMO) whose structure is presented in Fig.~\ref{Fig4:PCMO_TDep}a. Praseodymium (Pr)-based manganites have garnered much interest due to their tendency to display ordering phenomena relating to magnetic, orbital and charge degrees of freedom\cite{Murakami1998,Tomioka2001}, all of which can be controlled by chemical doping, magnetic fields and light\cite{Zhang_2016,Beaud2009}. The latter suggests that light can couple directly to critical infrared-active modes in the solid, potentially steering the system along pathways conducive to ultrafast phase switching \cite{Rini2007}. This mode-selective coherent control and the mechanism through which it is accomplished remains an active field of research whose viability can be assessed against a disordered phase transition scenario\cite{Wall2018,JohnsonPastor2024,Monti2022} using symmetry-resolved methods.     

Below 460 K, PCMO undergoes a structural phase transition from a high-temperature tetragonal state (HTT) to a low-temperature orthorhombic (LTO) distorted phase, driven by Jahn-Teller distortions that split the lattice parameters and break the structural point group symmetry (\ce{C4} $\rightarrow$ \ce{C2}) \cite{Salamon2001,Porer2020}. At 320 K, PCMO undergoes a metal-insulator transition due to charge and orbital ordering (COO), which induce a doubling of the unit cell along the \textbf{b} axis. With further lowering of the temperature, the distortions intensify causing lattice constants \textbf{a} and \textbf{b} to increasingly diverge from each other, until the onset of an antiferromagnetic phase below 120 K \cite{Chi2007}. Notably, the system retains in-plane anisotropy across these transitions, but it's magnitude varies\textemdash a behavior that can be sensitively monitored by measuring birefringence using linear RA spectroscopy. 

Using probe light at 1.3 \textmu m with horizontal polarization and an incident fluence 0.05 mJ cm$^{-2}$, we measured the birefringence of 001-orientated single crystal PCMO in the $ab$-plane to track changes in \ce{C2} symmetry as a function of temperature and optical pumping. The static results are presented in Fig.~\ref{Fig4:PCMO_TDep}, showing the measured temperature dependence between 200 K and 360 K. At 200 K, PCMO displays strong anisotropy, as seen in Fig.~\ref{Fig4:PCMO_TDep}b plotting $I_{\text{sig}}$, which has been fit to the expression derived previously in Eq.~(\ref{Eq:1}) showing excellent agreement with our birefringence model.

Figure~\ref{Fig4:PCMO_TDep}c provides an overview of the changes presented as polar plots, highlighting temperature-induced changes in \ce{C2} symmetry. As can be seen, upon increasing the temperature, the amplitude of anisotropy decreases (the radius is reduced) and is significantly suppressed at 360 K. However, the birefringence does not fully vanish, i.e. \ce{C4} symmetry is not restored. Additionally, the patterns undergo a detectable rotation at higher temperatures relative to the 200 K case (see also the inset of panel (d)). This effect can point to the formation of a different crystal axis or inversion of domains at higher temperatures. Crucially, such rotational changes are resolved only by dense angular sampling, as demonstrated here; normally, this would be missed in a measurement that records only two orthogonal angles \cite{singhla2013}. It should be noted that layered manganite compounds are known to form large single crystal domains, typically spanning 50–300 \textmu m\cite{Lee_2006}, and thus their formation is not expected to influence our measurements. This was confirmed by spatial scans around the probed region, which revealed a homogeneous response, with no detectable pattern rotation or amplitude variation.

Using Eq.~(6), we computed the temperature dependence of the order parameter, $\eta^2$, from the fitted DC and anisotropic terms. As shown in Fig.~\ref{Fig4:PCMO_TDep}d, a sharp change in $\eta^2$ occurs near 330 K, marking the COO transition temperature (\ce{T}$_{\mathrm{COO}}$). This transition, and the overall observed temperature dependence, is strongly corroborated by the XRD data of PCMO from Ref.~\!\citenum{Porer2020}, in which the authors measured the (2, $\tfrac{3}{2}$, 0) Bragg peak, corresponding to the COO state. Furthermore, the observed tilting of the crystallographic axis in our birefringence data offers valuable additional insight\textemdash information typically inaccessible via conventional optical spectroscopy and generally requiring full structural refinement from diffraction measurements. The excellent agreement between our optical results and reported diffraction data highlights the sensitivity and efficacy of our approach in tracking symmetry changes in quantum materials.

Having characterized the static response of PCMO to temperature, we subsequently excited the system to probe symmetry changes in the time domain. Near-IR (800 nm) pump pulses were employed at two fluence levels and three different repetition rates, and the PCMO sample was held at 300 K for these measurements. The temporal evolution of the order parameter and the change in reflectivity were tracked as a function of pump-probe delays by extracting the corresponding $\eta^2$ and $r_1^2$ parameters, respectively, at each time step, as outlined above.
\begin{figure*}[t!]
    \centering
    \includegraphics[width=17.8cm]{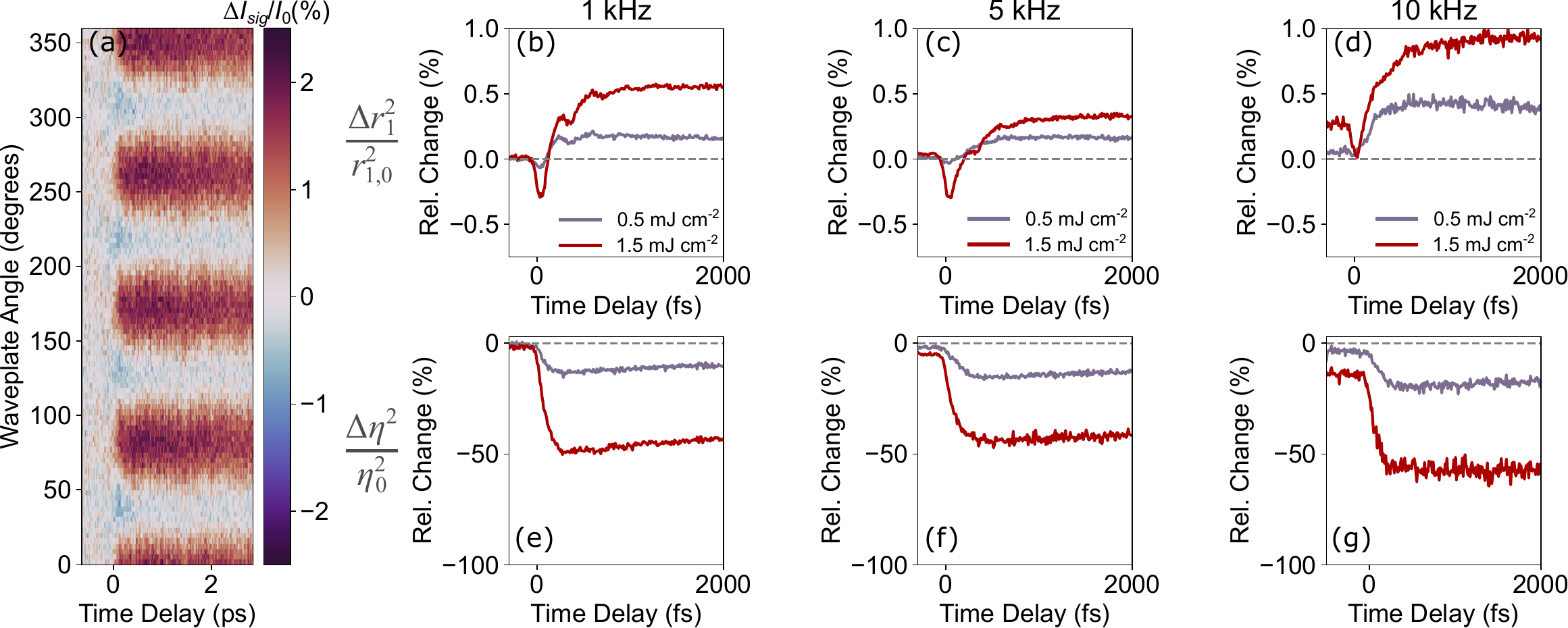}
    \caption{(a). Typical 2D tr-RA map of PCMO at 300 K  plotting intensity changes as function of time and angle following excitation using 800 nm pump (b-d). relative changes in reflectivity for two different fluences and three repetition rates (e-g). Corresponding changes in the order parameter as function of fluence and repetition rate.}  
    \label{Fig5:PCMO_TR}
\end{figure*}

Figure~\ref{Fig5:PCMO_TR}a shows a typical 2D time-resolved rotational anisotropy (tr-RA) map obtained from pump-probe experiments on PCMO. Upon photoexcitation, PCMO undergoes rapid changes in both lattice and electronic subsystems, as reflected in the symmetry and reflectivity responses, i.e. $\Delta \eta^2/\eta^2_0$ and $\Delta r^2_1/r^2_{1,0}$, respectively, where $\eta^2_0$ and $r^2_{1,0}$ represent the unpumped reference signals used for normalization. The corresponding results are presented in Fig.~\ref{Fig5:PCMO_TR}(b–d) and (e–g). Some notable observations can be readily made from these plots. Measurements at 10 kHz display a clear, strong pre-time zero signal: At 1.5 mJ cm$^{-2}$, the negative delay reflectivity signal, $\Delta r^2_1/r^2_{1,0}$, goes up by roughly 0.3\%, whereas the change in the order parameter, $\Delta \eta^2/\eta^2_0$ reaches approximately 15\%. Changes are predictably smaller at the lower fluence of 0.5 mJ cm$^{-2}$, but are not absent. The situation improves at lower repetition rates of 5 kHz and 1 kHz: the reflectivity data exhibit only minor changes before time zero at 5 kHz, high fluence and no change in the lower fluence case. Meanwhile, the corresponding pre-time zero order parameter signal is at least a factor of three smaller than at 10 kHz. The 1 kHz data, on the other hand, exhibit virtually zero offset in either parameter, at both fluence levels.

\begin{figure}[b!]
    \centering
    \includegraphics[width=8.0cm]{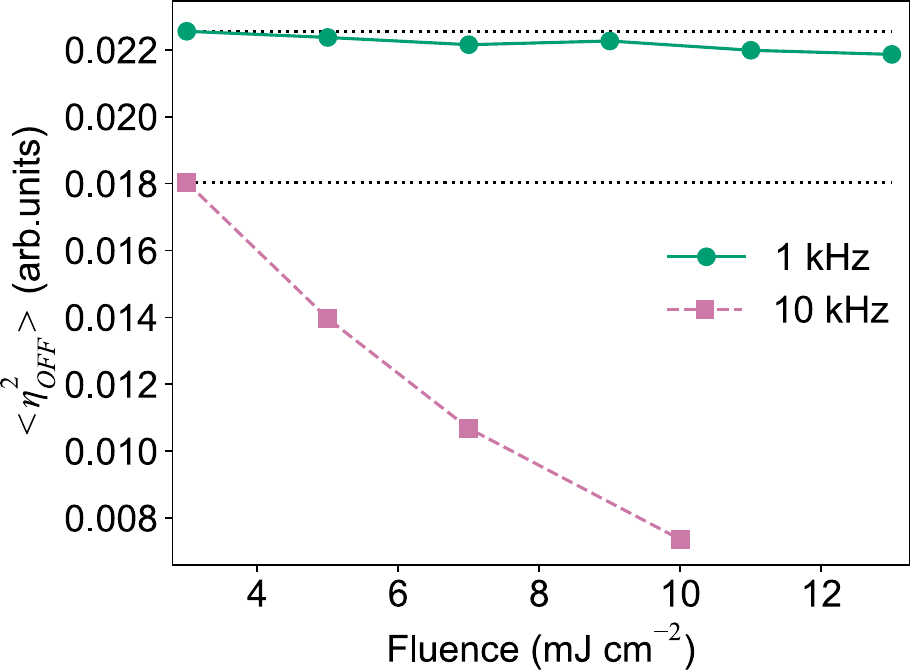}
    \caption{Plot showing the effect of repetition rate on the averaged OFF-channel signal in the presence of cumulative heating effects. The rate of change of signal against fluence is higher at 10 kHz (squares, dashed line) than at 1 kHz (circles, solid lines). The horizontal dotted lines serve as visual guides.}  
    \label{Fig6:PCMO_TR_off}
\end{figure}

A non-zero signal preceding the onset of dynamics (i.e. signal appearing at $\Delta t<0$) suggests an incomplete thermal recovery between laser pump shots, resulting in cumulative heating, and consequently, elevated base temperature\textemdash an effect that becomes more pronounced at higher repetition rates \cite{Vidas2020}. This behavior is more clearly illustrated by inspecting the OFF-channel (probe-only reference) $\eta^2 $ signal. Plotting the average OFF signal, $\langle\eta^2_{\mathrm{OFF}}\rangle$ against the fluence (Fig.~\ref{Fig6:PCMO_TR_off}) reveals a steeper gradient at 10 kHz compared to the much shallower slope at 1 kHz, confirming that\textemdash in the presence of cumulative heating\textemdash the rate of change of the OFF-channel signal is higher with fluence at higher repetition rates. This demonstrates an obvious but crucial point: while higher repetition rates are generally preferred for improving S/N and reducing acquisition times, the choice must consider the material system and the experimental conditions, including sample mounting, thermal conductivity, base temperature, etc. Such controls are crucial for guiding the selection of experimental parameters that ensure robust and reproducible results. In the present context, repetition rate emerges as a critical factor, and the ability to tune it, as enabled by our setup, further underscores its versatility.

\begin{figure*}[t!]
\centering
\includegraphics[width=17.5cm]{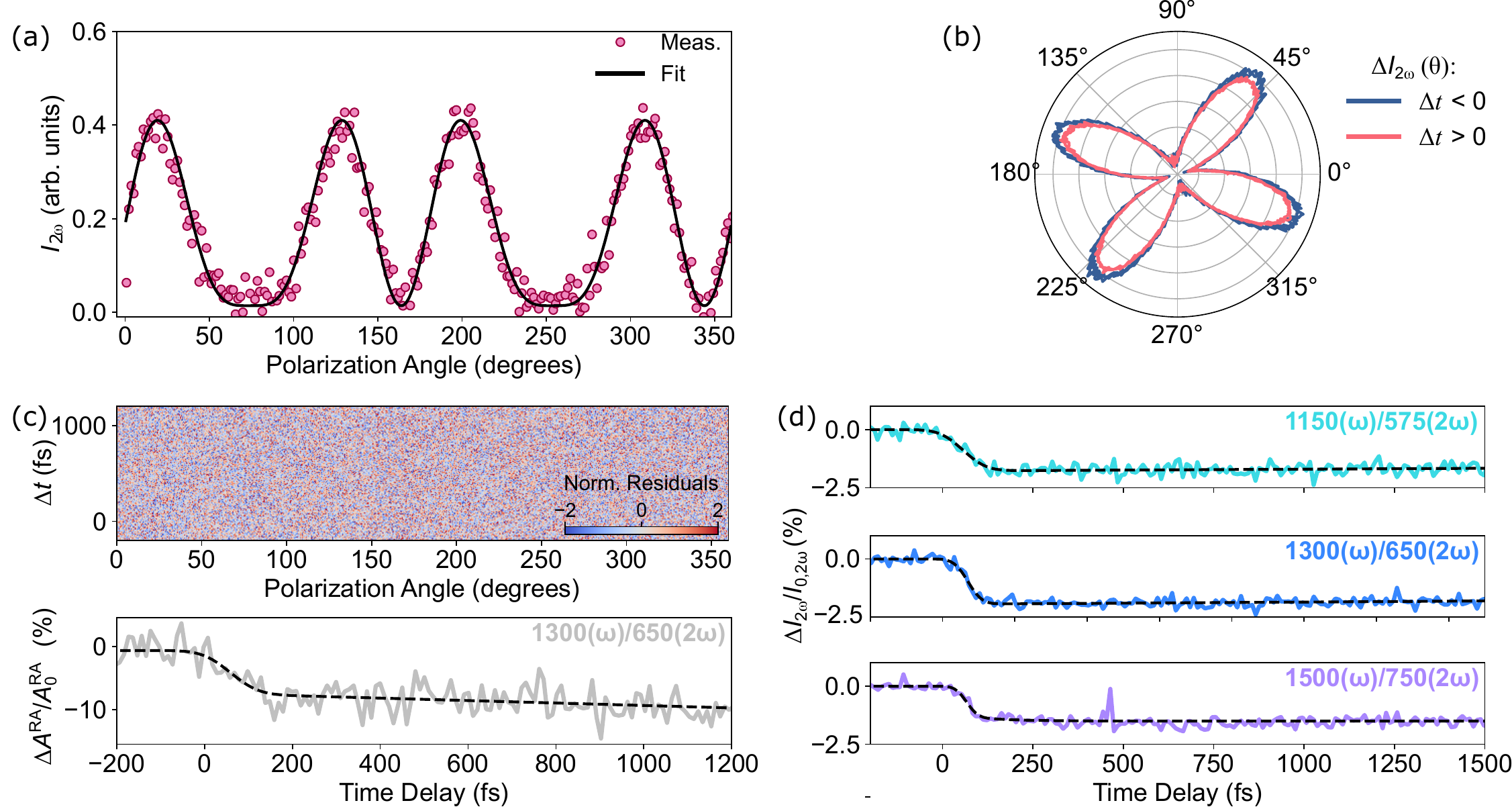}
\caption{(a). Measured SHG pattern from the GaAs crystal shows a four-lobed pattern which agrees with a (011) sample, as confirmed by the fit to Eq.~(8) (b). Polar representation of the average change to SHG-RA patterns before ($\Delta t<0$) and after ($\Delta t>0$) photoexcitation (c). \textbf{Upper}: A 2D map of the normalized residuals between the measured data and the model fit. The structureless map suggests that no photoinduced symmetry change occurred \textbf{Lower:} Extracted relative change in the amplitude, $A^{RA}$, by fitting the SHG-RA data, and (d). Time-dependent SHG intensity plots at a fixed $\theta$ angle for three different wavelengths given in units of nanometers, showing a reduction in the signal. The dashed lines are guides to the eye. }
\label{Fig7:SHG-RA}
\end{figure*}

It is also evident that the $\Delta \eta^2/\eta^2_0$ response is much more sensitive than $\Delta r^2_1/r^2_{1,0}$ in detecting pre-time zero signatures and also in discerning the nature of ultrafast phase transitions. In all cases, the reflectivity goes down immediately following photoexcitation and recovers partially within sub-500 femtoseconds before becoming constant. We also observe a 2.7 THz phonon oscillation in the signal, particularly in the 1 kHz data; the signal strength of the phonon becomes weaker with higher base temperature caused by stronger cumulative heating at 5 kHz and 10 kHz. By contrast, the $\Delta \eta^2/\eta^2_0$ response displays a drop without any notable recovery for the fluences employed here. This suggests that the \ce{C2} symmetry is rapidly disrupted due to photoexcitation, but the restoration takes several picoseconds. Finally, the absence of a coherent signal is also noteworthy, which implies a disorder-driven process, in contrast to the information captured in the $\Delta r^2_1/r^2_{1,0}$ signal. Indeed, the 2.7 THz oscillation has previously been attributed to the $A_{1g}$ Raman-active spectator mode of Ca/Pr ions \cite{Amelitchev2001}, which can be coherently launched but do not contribute to the phase transition. Usually, such insights on the nature of ultrafast phase transitions are obtained by challenging techniques, e.g. x-ray diffuse scattering experiments \cite{Wall2018,Monti2022}. In contrast, here we demonstrate that such information can also be extracted using symmetry-resolved optical experiments using a table-top setup. 

\subsection{Second harmonic generation rotational anisotropy}

Thus far, symmetry changes have been tracked using RA spectroscopy by measuring the linear response of materials using fundamental light. We have established that this approach yields invaluable information that may be difficult to disentangle or missing from conventional pump-probe optical spectroscopy. However, linear optical probes have a major limitation when applied to quantum materials in that they are insensitive to hidden quantum orders. For example, the antiferromagnetic (AFM) order is, in general, undetectable in the linear response because of the vanishing magnetic moment, rendering linear optics ineffective for studying many magnetic phase transitions \cite{Cheong2020}. Nonlinear probes, by contrast, can overcome this limitation and permit studies of hidden orders through their coupling with induced nonlinear polarization \cite{Fiebig2005,Denev2011,Zhao2018}. The simplest form of a nonlinear process is the second harmonic generation in which the frequency of an optical field, $\omega$, is doubled (i.e. 2$\omega$). The process is described by a second-order susceptibility tensor, $\chi^{(2)}$, whose non-zero, independent elements encode the information on magnetic, electronic and lattice structure\cite{boyd2008}. Thus, it becomes possible to identify crystallographic and electronic point group symmetries, by measuring different combinations of incident and emitted light polarizations. Additionally, SHG serves as a powerful tool for the sensitive probing of processes occurring at surfaces and interfaces\cite{Orenstein2021,Bergfeld2004,Fiebig2023}. 

We incorporated the option to measure the second-order nonlinear response of materials in our setup. To this end, a simple and straightforward change was made: the $\mathrm{PD_{sig}}$ detector was replaced by an avalanche photodiode (APD) to make it more sensitive to measure extremely weak signals, as expected from second harmonic processes. Furthermore, bandpass filters (with central wavelength at the second harmonic, $\lambda_{2\omega}$ and bandwidth = 40 nm FWHM) of combined optical density (OD) > 11 were placed in front of the APD to suppress the fundamental light at $\omega$ and only allow 2$\omega$ signals to be recorded. Finally, recall that the ultrabroadband HWP used in linear optical experiments also enables SHG-RA measurements without requiring exchange or realignment.

\subsection*{Second harmonic generation in gallium arsenide}
We carried out SHG-RA experiments on a single crystal of commercially available gallium arsenide (GaAs) with a (011) surface normal. Gallium arsenide is a III-V semiconductor with a direct band gap of $\approx$ 1.4 eV and possesses a zinc blende crystal structure. More specifically, it is characterized by the non-centrosymmetric $\text{T}_d$ point group and exhibits a high nonlinearity coefficient, making it ideal for second harmonic generation. The $\chi^{(2)}$ for $\text{T}_d$ point group has the following non-vanishing tensor elements:  $xyz=xzy=yzx=yxz=zxy=zyx$, where $x$, $y$, and $z$ denote the crystallographic axes. Assuming a parallel-parallel detection, the following functional form is derived for the intensity, $I^{PP}_{2\omega}$, from the 011 sample with respect to the incident polarization angle, $\theta$\cite{Zu2022,Sanger2006},

\begin{align}
        % I^{PP}_{2\omega}(011) = \frac{9A\sin^4\Phi\cos^2\Phi}{4}
        I^{pp}_{2\omega}(\theta) = \frac{9A}{64} \left[ 1 - \frac{1}{2} \cos(2\theta) - \cos(4\theta) + \frac{1}{2} \cos(6\theta) \right]
\label{Eq:GaAs}.
\end{align}

where \textit{A} is the amplitude. We used the 1.3 \textmu m output of the OPA as a probe for our measurements. The fluence was adjusted to 1 mJ cm$^{-2}$, and the laser repetition rate was set to 10 kHz. The incident polarization was rotated at twice the rate of the HWP spinning at 10 Hz, and the SHG light at 650 nm was filtered from the fundamental and detected by the APD as a function of polarization rotation. Figure~\ref{Fig7:SHG-RA}a shows the SHG-RA signal from GaAs (011) plotted against the polarization angle, $\theta$, averaged over ten sweeps. We obtain a pattern consistent with the expected four-fold symmetry of GaAs (011)\cite{Sanger2006}, as confirmed by the agreement with the fit using Eq.~(\ref{Eq:GaAs}) for the (011) surface, validating our results. We reiterate that a single pattern corresponding to a complete rotational cycle takes 50 ms and clearly exposes the lobed features, even at low probe intensities, thanks to the high achievable S/N.   

Having demonstrated the acquisition of high-quality static SHG-RA data, we proceeded to measure the pump-probe response to photoexcitation. Using fundamental light emitted by our laser system at 800 nm as the pump, we excited the GaAs (011) sample with a fluence of 5 mJ cm$^{-2}$. Full SHG-RA patterns were recorded at each step in the pump-probe delay, spanning up to 1.2 ps. Figure~\ref{Fig7:SHG-RA}b summarizes the changes in the signal by comparing the average SHG intensity measured before, $\Delta I_{2\omega} (\theta, \Delta t<0)$, and after, $\Delta I_{2\omega} (\theta, \Delta t>0)$, photoexcitation. It can been seen that photoexcitation at 800 nm modulates the SHG response, leading to an overall reduction in the detected intensity. Using Eq.~(\ref{Eq:GaAs}), we modeled the time-dependent SHG-RA data. Our model assumes that the intrinsic $\chi^{(2)}$ tensor components of GaAs remain constant in their relative values and that photoexcitation only scales the overall signal amplitude, $A$. First, we verified whether this assumption was justified and if any photoinduced symmetry change occurred. If the crystal symmetry changed upon excitation, it could be reflected in the normalized residuals (residuals divided by the standard deviation) as systematic, unfitted angular features. In our fits, $A$ and a baseline offset, \textit{C}, were the only two free parameters.

In the upper panel of Fig.~\ref{Fig7:SHG-RA}c, we present the results of the residual analysis of the measured SHG-RA signal in the form of a 2D map of normalized residuals of the full dataset. The figure provides no indication of an induced symmetry change in our data, as evidenced by the featureless map, and further shows that the model in Eq.~\ref{Eq:GaAs} captures the data well. We note that the purpose here is not to present a rigorous statistical analysis of the data, but to demonstrate the setup functionalities and the consistency of the fits. We extracted the time-dependent changes in the fitted amplitude from the SHG-RA measurements, i.e. $\Delta A^{RA}/A_0^{RA}$, where $A_0^{RA}$ denotes the unpumped reference. The plot of relative changes in $A^{RA}$, provided in Fig.~\ref{Fig7:SHG-RA}c (lower panel), reveals that the SHG amplitude is suppressed by $\approx10$\% following photoexcitation without a hint of recovery within the measurement time window. 

Additionally, we performed time-resolved SHG spectroscopy experiments at a fixed $\theta$ to record the corresponding changes in the SHG intensity ($\Delta I_{2\omega}/I_{0,2\omega}$, where the denominator represents the reference SHG intensity in the OFF-channel) for three different probe wavelengths. The fluence of the 800 nm pump beam was lower in this case than in the measurements described above, at 1.25 mJ cm$^{-2}$. The results, presented in Fig.~\ref{Fig7:SHG-RA}d, exhibit similar kinetics across the probe wavelengths used. Together, both sets of measurements indicate that the SHG signal is reduced when pumping at 800 nm, presumably due to the generation of carriers leading to photoinduced heating or screening of the dipole moment, thereby making the SHG process less efficient, with a weak dependence on the probe color. This latter observation implies that the suppression of SHG intensity is likely not due to a resonance shift \cite{Luppi2010}, but rather that the entire dipole moment is affected by photoexcitation. Regardless of the true underlying cause of this effect, the results demonstrate the feasibility of incorporating time-resolved SHG experiments with pump excitation in our setup, enabling both full RA measurements and those at fixed angles.

\subsection*{Single-photon detection: a path towards SHG-RA in quantum materials}
\label{section:SPD}
The example above represents the best-case scenario for recording $2\omega$ photons because of the high nonlinearity of GaAs. In contrast, quantum materials present a more stringent challenge for our setup due to their orders of magnitude weaker SHG signals, which precludes the use of the APD. Therefore, we turn to single-photon detection schemes. Unlike SHG setups which require high incident powers, high repetition rates, photomultiplier tubes or lock-in amplifiers, we propose a more simple approach of using a single-photon detector (SPD) operating in the Geiger mode and gating it to suppress background. In this section, we describe this method and show that it allows reducing the incident flux by more than an order of magnitude\textemdash a critical consideration in the studies of quantum materials\cite{Zhang_2016}\textemdash compared to other setups and still achieves a high count rate of SHG photons. To characterize the SPD response, we utilized a different setup from that described above. In this configuration, the beam passed through a rotating half-wave plate (HWP), impinged on the sample, and the outgoing beam was analyzed using a polarizer, i.e. a more conventional scheme that did not involve a double-pass through the HWP was employed at near-normal incidence. This setup allowed us to directly compare our results with those reported in the literature. 
\begin{figure}[t!]
    \centering
    \includegraphics[width=8.65cm]{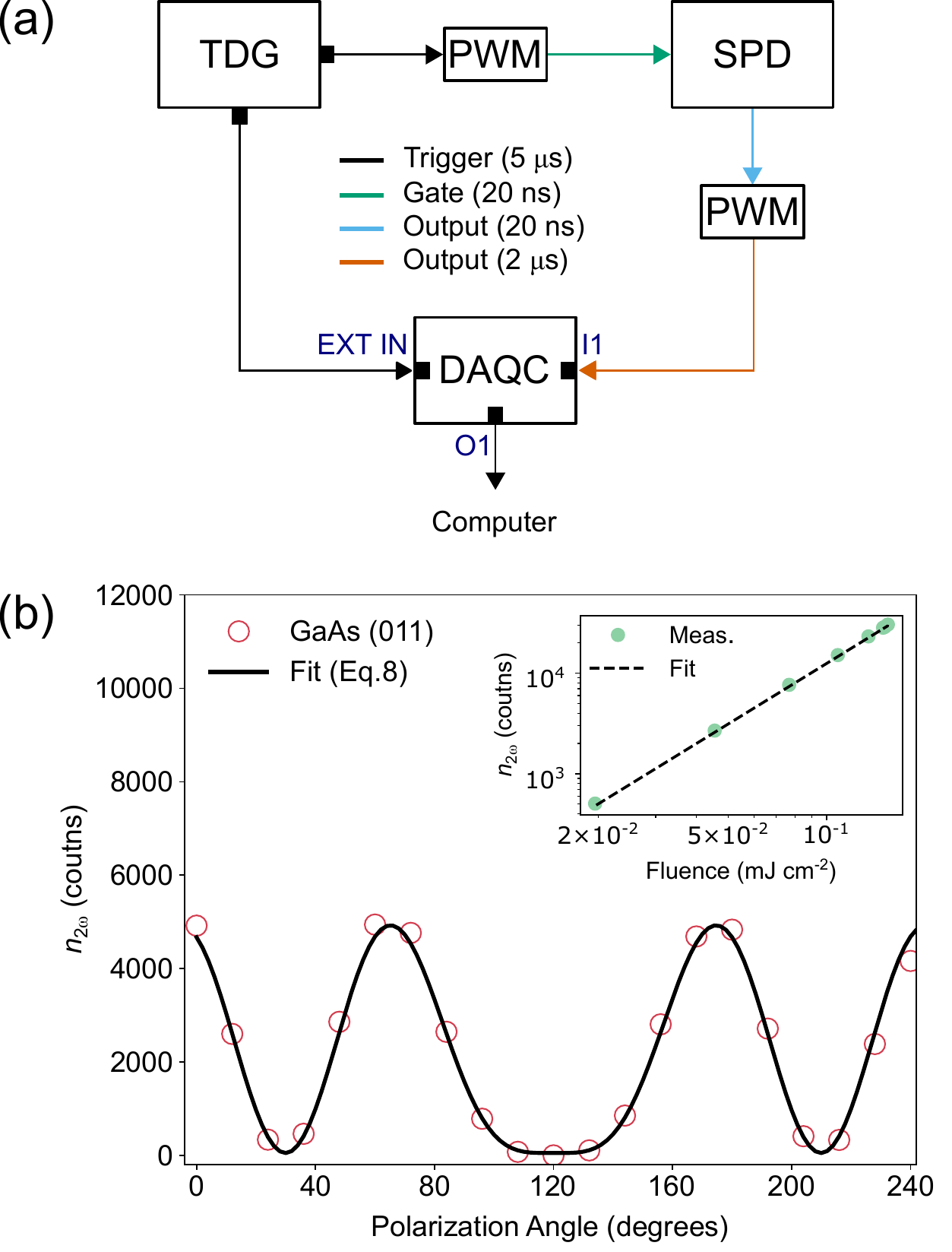}
    \caption{(a). Circuit diagram for the SPD setup. TDG: time delay generator; PWM: pulse width modulator; SPD: single photon detector; DAQC: data acquisition card; EXT IN: external input. O1,2 are the output channels of the DAQC and I1 is one of the inputs of the DACQ (b). Measurement of SHG from GaAs(011) crystal surface using SPD. Inset shows the photon counts as a function of laser fluence in a log-log format. The resulting straight line establishes a quadratic relationship between detected SHG photons and the incident flux.  }
    \label{Fig8:SPD}
\end{figure}

Figure~\ref{Fig8:SPD}a shows a circuit diagram that illustrates the approach. A trigger TTL pulse from the laser of five microseconds duration is first modulated by a home-built pulse width modulator (PWM) to produce a twenty nanoseconds pulse used to trigger/gate the SPD. This pulse defines the detection window in which all photons impinging on the detector are recorded. Following the acquisition event, the SPD outputs pulses of the same duration as the original gate (i.e. 20 ns), which are then stretched by a second PWM unit to $\approx 2$ \textmu s to meet the sampling rate requirement of our data acquisition card (DAQC). The digitized data are subsequently transferred to a computer for processing.

To evaluate the effectiveness of this approach, we once again measured the signal from GaAs (011) using photons at 1300 nm. However, compared to earlier measurements, the incident fluence was reduced by a factor of twenty to 0.05 mJ cm$^{-2}$ (a two-orders-of-magnitude reduction from a previous SHG-RA work on GaAs using 800 nm probe, as reported in Ref.~\!\citenum{Harter2015}). The data were obtained by stepping the polarization angle and exposing the SPD for ninety seconds at each measured angle in a parallel-parallel polarization geometry. The resulting SHG-RA pattern is provided in Fig.~\ref{Fig8:SPD}b and reproduces the characteristic shape observed for GaAs (011) in Fig.~\ref{Fig7:SHG-RA}a. Furthermore, to confirm that SPD predominantly detected SHG photons, we performed a power dependence calibration, which yielded a quadratic relationship between the measured SHG photon counts ($n_{2\omega}$) and the incident flux (fluence); as shown in the inset of Fig.~\ref{Fig8:SPD}b, the log-–log plot displays a linear trend, confirming the second-order nature of the signal. These findings substantiate the robustness of our approach in sensitively detecting SHG signals.   
\begin{figure}[t!]
    \centering
        \includegraphics[width=8.65cm]{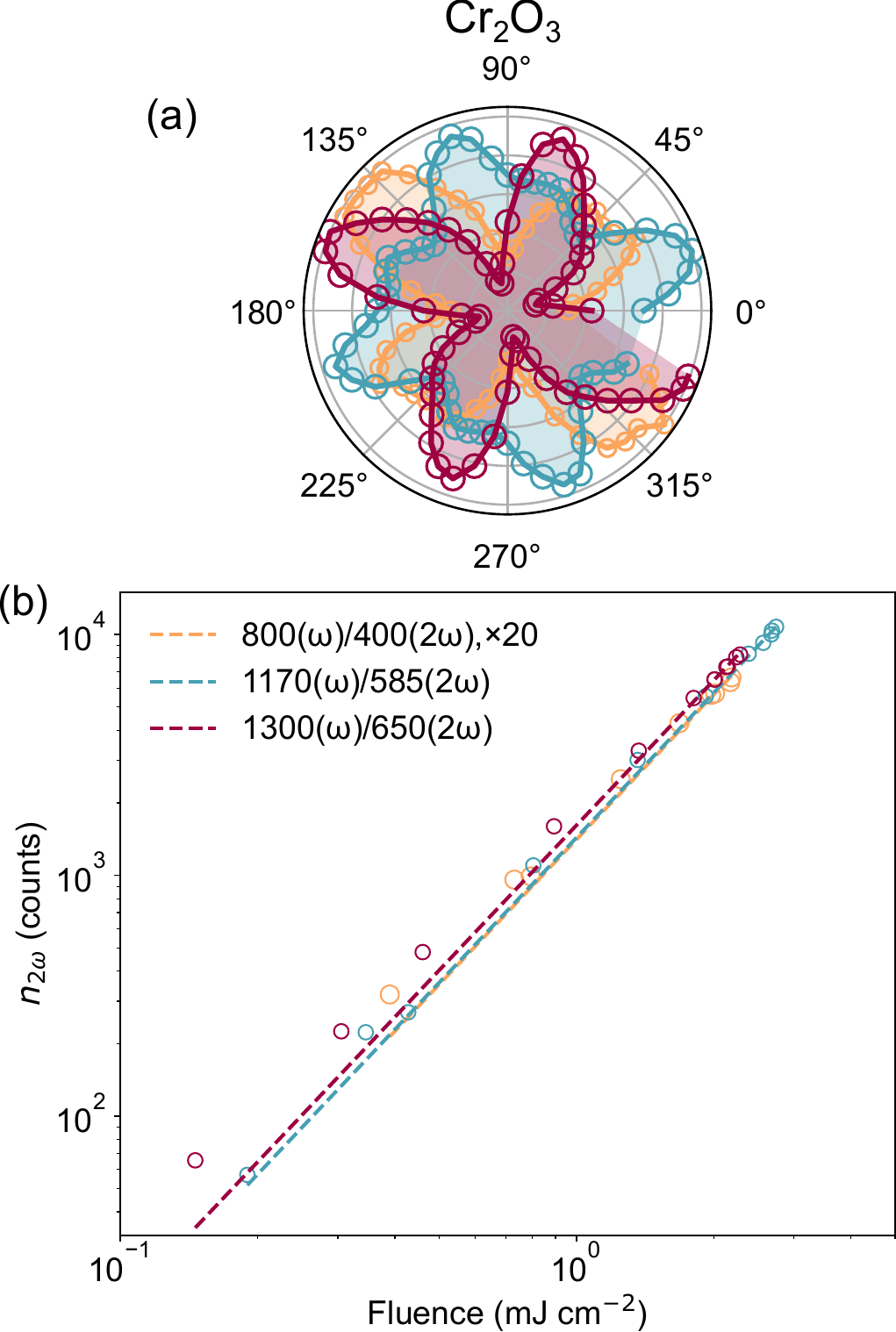}
    \caption{(a). Second harmonic generation from \ce{Cr2O3} at three different wavelengths. Different anisotropy patterns are obtained based on the probe color. Here, the SHG intensities have been normalized for the sake of comparison (b). The quadratic dependence observed in the log--log plot, represented as a straight line, confirms the detection of SHG photons. }
    \label{Fig9:CROPol}
\end{figure}

After benchmarking against the GaAs SHG signal, we investigated chromium (III) oxide (\ce{Cr2O3}), a well-studied antiferromagnetic, insulating quantum material in which AFM ordering breaks time-reversal and inversion symmetries below its N{\'e}el temperature, $T_N$, of 307 K \cite{Birss1966,Zhang2024}. Therefore, \ce{Cr2O3} gives rise to electric dipole allowed SHG below $T_{N}$ enabled by the polar source term. Conversely, above $T_{N}$, \ce{Cr2O3} possesses $\bar{3}m$ symmetry and is centrosymmetric. The polar term thus vanishes, but an axial source term is allowed, giving rise to magnetic dipole SHG. We carried out measurements of a commercially obtained single crystal of \ce{Cr2O3} at 300 K. At this temperature, the efficiency of SHG is expected to be very small and, thus can additionally serve to test the limits of our detector. The polarization angle was once again stepped, and the data were recorded for a maximum exposure of thirty seconds. 

Figure~\ref{Fig9:CROPol} shows the results of polarization dependence of SHG from \ce{Cr2O3} for three different probe wavelengths. The obtained quadratic dependence (again seen as straight lines in log--log plots, see Fig.~\ref{Fig9:CROPol}b) provides confidence in the detection of photons at the second harmonic of each probe wavelength (585 nm, 650 nm and 400 nm). Moreover, the results showcase the effect of different probe wavelengths on the resulting SHG-RA patterns due to wavelength-dependent nonlinear susceptibility\cite{Fiebig1994}: at 1170 nm, the highest count rate of SHG photons is obtained due to resonance enhancement, while the efficiency is lowest for 800 nm. In a previous study, Fiebig and co-workers measured the SHG spectrum up to 3.1 eV (i.e. $2\omega = 400$ nm), offering a new approach to imaging domain topography in magnetic materials, although strong contrast at this energy was not reported.\cite{Fiebig1995}. In our case, SHG measurements at $\omega = 800$ nm reveal a clear polarization dependence which demonstrates the high sensitivity achieved with our detection scheme. Furthermore, the wavelength-dependent evolution of SHG-RA patterns observed here can help uncover useful details, such as the relative contributions of axial (magnetic) and polar (electric) source terms and underscores the importance of spectrally-resolved SHG measurements for disentangling these contributions.
\begin{figure}[t!]
    \centering
    \includegraphics[width=8.65cm]{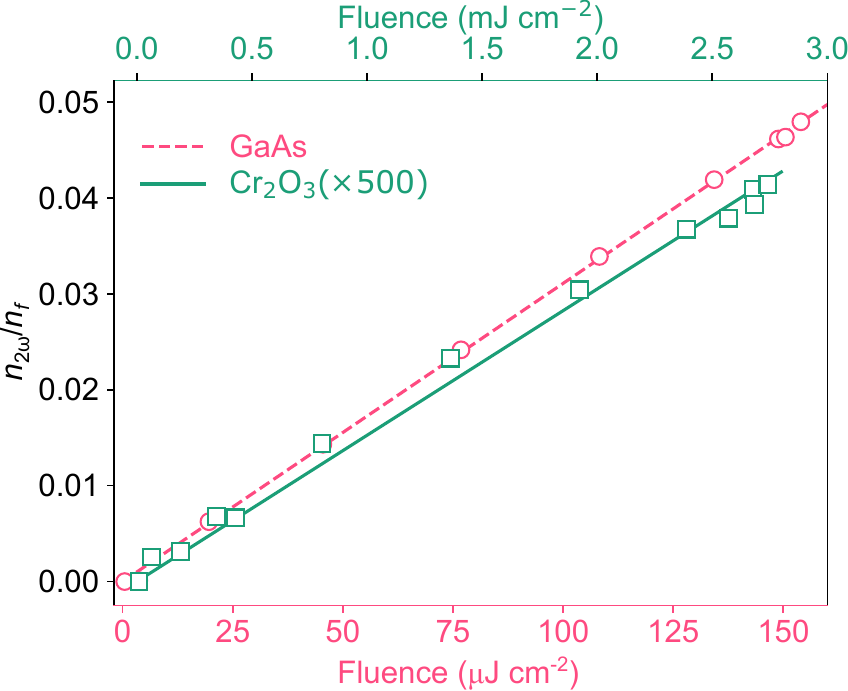}
    \caption{Comparison of SHG intensity between GaAs (circles, dashed line) and \ce{Cr2O3} (diamonds, solid line). Upper fluence axis (GaAs) and lower fluence axis (\ce{Cr2O3}). The signal for \ce{Cr2O3} has been multiplied 500 times. Note that the fluence axes differ by more than an order of magnitude.}
    \label{Fig10:GaAsVsCRO}
\end{figure}

Finally, comparing GaAs and \ce{Cr2O3} in terms of the ratio of measured SHG photons to the number of incident fundamental photons, i.e. $n_{2\omega}/n_{f}$, shown in Fig.~\ref{Fig10:GaAsVsCRO}, it can be seen that \ce{Cr2O3} displays at least two-orders-of-magnitude weaker SHG signal compared to GaAs at this temperature. Despite this, we could confidently measure SHG-RA patterns with high fidelity and S/N. This impressive gain in sensitivity opens up the path for future time-resolved SHG experiments on different quantum materials in our setup using SPD, incorporating recording of full rotational sweeps within a single shot.

\section{Outlook and Conclusions} 
\label{section:conclusions}
In summary, we have presented a new setup designed to carry out static and time-resolved linear and SHG rotational anisotropy experiments on quantum materials in the optical regime. Our method provides the capability to record population dynamics and enables simultaneous tracking of the symmetry changes occurring on femtosecond timescales. The ability to easily adjust the repetition rate, probe wavelengths, sample temperature, and measure both linear and nonlinear response allows numerous ways to explore material behavior distinguishing our setup from many reported in the literature. 

Although we have demonstrated the core functionality of the setup in this paper, there are several planned future improvements. One key area is enhancing the time resolution of the setup, which is currently limited by transmissive optics (e.g. thick achromatic HWP). We are investigating careful chirp control of the pump/probe pulses as a promising solution to mitigate temporal smearing. This would extend the capability of the setup to measure phonons with frequencies exceeding 7 THz (e.g. relevant in perovskite quantum materials\cite{singhla2013, Amuah2021}) and even resolve higher harmonic modes.

While the fundamental output of Ti:Sa laser at 800 nm is currently being used as the pump\textemdash suitable for many quantum materials with band gaps in the mid-IR and THz range\textemdash, expanding the laser's capabilities to the visible range would allow for a broader pallet of samples to be studied, such as 2D semiconductors and magnetic materials. As mentioned earlier, the flexibility to swap the pump and probe arms on demand, along with frequency conversion (e.g. SHG/THG) of the fundamental beam and the OPA output, can facilitate this option.

Another potential improvement involves the ability to measure different input-output polarization combinations. Currently, only parallel-parallel geometry is implemented, which can limit the exploration of nonlinear susceptibility tensor elements, especially in SHG experiments\cite{Morey2025}. A straightforward solution is to replace the WGP in the setup above with a cube polarizer and split the returning beam into parallel and perpendicular components. This would require an additional photodiode and an extra channel on the DAQ system, especially if parallel and orthogonal channels are to be measured simultaneously. 

The noninvasive design of the setup further allows more exotic experiments to be carried out, such as the integration of replica pump pulses with an adjustable delay between them, enabling the investigation of phenomena such as double pumping and its impact on structural dynamics \cite{JohnsonPastor2024}. It also opens the avenue to study inversion symmetry-breaking processes under high hydrostatic pressures with minimal modification of the original design. 

We believe that the aforementioned improvements will further enhance the scope of experiments that can be performed on the setup. The vast potential to explore the various parameters opens numerous possibilities, making for a versatile platform for symmetry-resolved dynamics on ultrafast timescales in quantum materials and beyond. These capabilities pave the way for studying a wide array of novel phenomena in thin films and bulk crystals.

\begin{acknowledgments}
\noindent This research was funded by the Danish Carlsbergfondet (Grant No. CF20-0169). N. Khatiwada acknowledges generous funding from Erasmus Mundus QuanTEEM program for studentship. The authors acknowledge the excellent support from AU physics mechanical and electronics workshops for their enormous support. Especially help from Anders Strøbech Damgaard, René Warthoe Davids, Torben Hyltoft Thomsen, Henrik Juul, Erik Loft Larsen, Frank Daugaard and Martin Stougaard is gratefully acknowledged.
    
\end{acknowledgments}

\section*{Conflict of interest}
\noindent The authors have no conflicts of interest to disclose.

\section*{Author Contributions}
\noindent \textbf{Khalid M. Siddiqui}: Conceptualization (equal); Data curation (equal); Formal analysis (equal); Investigation (equal); Methodology (equal); Writing – original draft (lead);  Writing – review \& editing (equal); \textbf{Hanna  Strojecka}: Methodology (equal); Formal analysis (equal); Investigation (equal); Writing – review \& editing (equal); \textbf{Thomas H. Meyland}: Methodology (equal); Formal analysis (equal); Investigation (equal); Writing – review \& editing (equal); \textbf{Nitesh Khatiwada}: Methodology (equal); Formal analysis (equal); Investigation (equal); Writing – review \& editing (equal); \textbf{Nikolaj Klinsby}: Methodology (equal); Investigation (equal); Writing – review \& editing (equal); \textbf{Daniel Perez-Salinas}: Methodology (equal); Formal analysis (equal); Investigation (equal); Writing – review \& editing (equal); \textbf{Simon E. Wall}: Conceptualization (lead); Data curation (equal); Methodology (equal); Formal analysis (equal); Investigation (equal); Writing – original draft (equal); Writing – review \& editing (equal); Funding acquisition (lead); Supervision (lead).

\section*{Data Availability Statement}
The data that support the findings of this study are available from the corresponding authors upon reasonable request.

\end{document}